\documentclass{aa}
\usepackage{psfig}

\newcommand{\degrees}{$^{\circ}$}
\newcommand{\etal}{et al.}

\newcommand{\NeII}{\mbox{Ne\,{\sc ii}}}

\newcommand{\ArII}{\mbox{Ar\,{\sc ii}}}

\newcommand{\mum}{\mbox{$\mu$m}}

\begin{document}
\thesaurus{08	
	(09.09.1 Ced\,201; 09.04.1; 13.09.4)}
\title {ISO observations of the reflection nebula Ced\,201:
evolution of carbonaceous dust
\thanks{Based on observations at the Cal Tech submillimeter observatory (CSO)
and with ISO, an ESA project with instruments funded by ESA
member states (especially the PI countries: France, Germany, the
Netherlands and the United Kingdom) and with the participation of ISAS
and NASA.}}

\author{D. Cesarsky\inst{1,2}\and
        J. Lequeux\inst{3}\and
	C. Ryter\inst{4}\and
	M. G\'erin\inst{5}
}
\offprints{james.lequeux@obspm.fr}
\institute{	
Institut d'Astrophysique Spatiale, Bat. 121, Universit\'e Paris
XI, F-91450 Orsay CEDEX, France
\and
Max Plank Institut f\"ur extraterrestrische Physik, Postfach 1603, 
D-85740 Garching, Germany
\and
DEMIRM,Observatoire de Paris, 61 Avenue de l'Observatoire, F-75014
Paris, France
\and
SAp/DAPNIA/DSM, CEA-Saclay, F-91191 Gif sur Yvette CEDEX, France
\and
Ecole Normale Sup\'erieure, 24 Rue Lhomond, F-75238 Paris CEDEX 05, France
}
\date{Received 16 November 1999; accepted 28 January 2000}
\maketitle 
 
\markboth{Cesarsky D., Lequeux J., Ryter C. et al.}
         {Ced\,201 and evolution of carbonaceous dust}

\begin{abstract}
We present spectrophotometric imaging mid--IR observations of the reflection
nebula Ced\,201. Ced\,201 is a part of a molecular cloud illuminated by a B9.5V
star moving through it at more than 12 km s$^{-1}$.  The spectra of 
Ced\,201 give evidence for transformation  of very small carbonaceous 
grains into the carriers of the Aromatic Infrared Bands (AIBs), 
due to the radiation field of the illuminating star and/or to shock 
waves created by its motion. These very small grains emit mainly very 
broad bands and a continuum. We suggest that they are present
everywhere in the interstellar medium but can only be detected in the
mid--IR under special circumstances such as those prevailing in this
reflection nebula. The efficiency of energy conversion of stellar light into
mid--infrared emission is 7.5\%~  for both the very small grains and
the AIB carriers, and the fraction of interstellar 
carbon locked in these emitters is approximately 15\%. 
\end{abstract}
\section{Introduction}
The mid--infrared emission bands at 3.3, 6.2, 7.7, 8.6, 11.3 and 12.7\,\mum~
are ubiquitous in the interstellar medium (ISM). The coincidence of these bands
with characteristic wavelengths of aromatic molecules and attached functional
groups (see e.g. Allamandola et al. \cite{Allamandola89}) is sufficiently 
convincing to call them Aromatic Infrared Bands (AIBs). Other, more  neutral
designations are Unidentified Infrared Bands (UIBs) or Infrared  Emission
Features (IEFs). The designation should stabilize in the future.  Their
carriers are very small grains (or big molecules) heated transiently by the
absorption of a single  UV (Sellgren et al. \cite{Sellgren83}) or visible
(Uchida et al.  \cite{Uchida98};  Pagani et al. \cite{Pagani99}) photon. They
are often identified  with Polycyclic Aromatic Hydrocarbons (PAHs) (L\'eger \&
Puget \cite{Leger}, Allamandola et al. \cite{Allamandola85}). Duley \& Williams
(\cite{Duley88}) and Duley et al. (\cite{Duley93}) have proposed an alternative
interpretation  in which the carriers of the AIBs are not free particles but
``graphitic  islands'' (i.e. akin to PAHs) loosely attached to each other so
that the heat  conductivity is small. Then their thermal behaviour is
comparable to that of  free particles. Similar models have been also proposed
by others. The carriers could also be \mbox{3--D} very small particles. In this
Letter, we will not enter  into this debate and will simply call them the {\it 
AIB carriers}.

The origin of the AIB carriers is still unclear. The idea that they are formed 
in the envelopes of carbon stars is not supported by observation and meets 
with theoretical difficulties (e.g. Cherchneff et al. \cite{Cherchneff}).  They
might be formed outside carbon star envelopes from  carbonaceous particles
condensed previously in these envelopes. Schnaiter et al. (\cite{Schnaiter99})
and others have proposed such an evolutionary scheme from carbon stars to 
planetary nebulae. However the AIB carriers formed in this way might not 
survive the strong UV field of planetary nebulae. For example, Cox et al. 
(\cite{Cox}) find no AIB in the Helix nebula, an old carbon--rich planetary 
nebula. Another possibility is that they are formed in the ISM from 
carbonaceous grains or carbonaceous mantles covering silicate nuclei. Boulanger
et al. (\cite{Boulanger}) and Gry et al. (\cite{Gry}) have suggested that   
the strong variations in the IRAS 12\,\mum/100\,\mum~ interstellar flux ratio
are due to the release of transiently heated particles (emitting at 12\,\mum)
by bigger particles, respectively due to  UV irradiation or to shattering by
shocks. Thanks to the Infrared Space Observatory (ISO), we know now that in the
general ISM the emission in the  IRAS 12\,\mum~ band is dominated by AIBs.
Variations in the ratio of AIB carriers to big grains are thus confirmed.
Laboratory experiments shed  some light  on this process. Scott et al.
(\cite{Scott97}) have observed the release of aromatic carbon clusters
containing in excess of 30 carbon atoms  by solid hydrogenated amorphous carbon
(HAC)  irradiated by a strong UV laser pulse. The energy deposited by the laser
pulse per unit target area is similar to the energy of a central  collision
between two grains of 10 nm radius at a velocity of 10 km s$^{-1}$. This is an
indication for a possible release of AIB carriers in interstellar  shocks.  

ISO has also shown the existence of variations of AIB spectra that may be
related to a transformation of carbonaceous grains.  Recently, Uchida et al.
(\cite{Uchida00}) have observed a broadening of the 6.2, 7.7 and 8.6\,\mum~
AIBs when going from strong to weak radiation field in the reflection nebula
vdB\,17 = NGC\,1333. They also notice an increase of the flux ratio
$I$(5.50--9.75\,\mum)/$I$(10.25--14.0\,\mum) with increasing radiation field in
several reflection nebulae. 

We present mid--IR spectrophotometric and CO(2--1) line observations of 
another reflection nebula, Ced\,201, that shed light into this transformation.
Section 2 describes briefly the observations and reductions. Sect. 3 describes
the results, and Sect. 4  contains a discussion and the conclusions. 

\section{ISO observations and data reduction}

The observations have been made with the 32$\times$32 element mid-infrared
camera (ISOCAM) on board of ISO, with the Circular Variable Filters (CVFs) (see
Cesarsky et al. \cite{CCesarsky} for a complete description). The observations
employed a 6\arcsec/pixel magnification, yielding a field of view of about
3\arcmin$\times$3\arcmin.  Full scans of the two CVFs in the long-wave channel
of the camera have been performed, covering a wavelength range from 5.15 to
17\,\mum. 10 exposures of 2.1s each were added for each step of the CVF, and 20
extra exposures were added at the beginning of each scan in order to limit the
effect of the transient response of the detectors.  The total observing time
was about 1 hour. The raw data were processed as described in Cesarsky
\etal~(\cite{M17}) using the CIA software\footnote{CIA is a joint development
by the ESA Astrophysics Division and the ISOCAM Consortium led by the ISOCAM
PI, C. Cesarsky.} (Starck et al. \cite{Starck}).  However, the new transient
correction described by Coulais \& Abergel (\cite{Coulais}) has been applied,
yielding considerable improvements with respect to previous methods.

\section{Results}
\begin{figure}[t!]
\vspace{-0.8cm}
\psfig{file=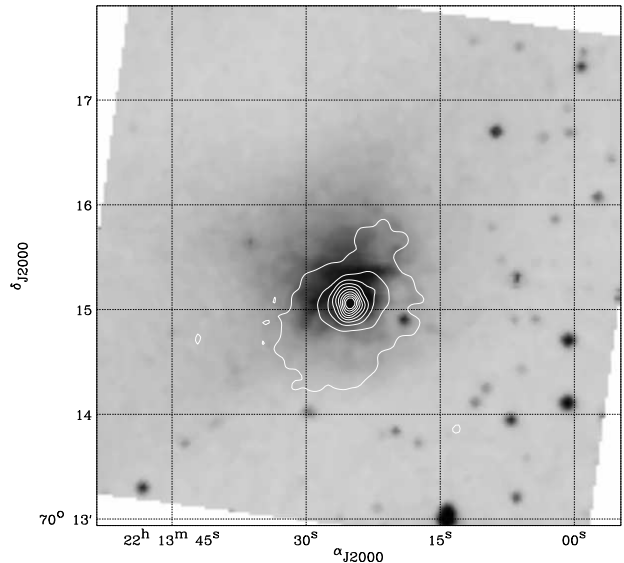,width=8.8cm,angle=0.0}

\caption{Contours of the total mid-IR emission (5--17\mum) of Ced\,201
superimposed over the Digital Sky Survey image. Note the absence of stars in 
the NE half of the image, due to extinction by the molecular cloud.}
\end{figure}

\begin{figure}[t!]
\psfig{file=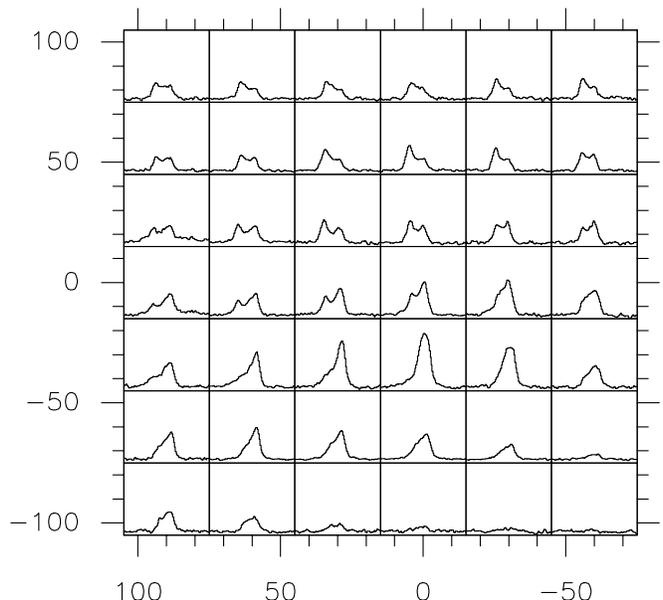,width=8.8cm,angle=270.0}
\caption{Spectra of Ced\,201 in the CO(2--1) line, obtained with the Caltech
Submillimeter Telescope (HPBW 30\arcsec) on a 30\arcsec$\times$30\arcsec 
grid. For each spectrum, abscissae give the LSR velocities from -8 to -2
km s$^{-1}$ and ordinates give the antenna temperatures $T_A^*$ from -1 to
20 K. The main--beam efficiency of the telescope is 0.76. The coordinates
are in arcsec relative to the reference 
position $\alpha$(J2000) = 22h 13m 24.4s, $\delta$(J2000) = 70\degrees~ 
15\arcmin~ 24\arcsec. The exciting star is 19\arcsec~ to the south. The CO 
line emission peaks in the direction of the star.
}
\end{figure}

The reflection nebula Ced\,201 is a rather compact object at a distance of 420
pc (Casey \cite{Casey}), on the edge of a molecular cloud. It is excited  by
the B9.5\,V star BD +69\degrees1231.   Witt et al. (\cite{Witt}) notice that
the radial velocity of this star differs from that of the molecular cloud by
11.7 $\pm$ 3.0 km s$^{-1}$   so that Ced\,201 is probably the result of an
accidental encounter of the star with the molecular cloud, while for most other
reflection nebulae the exciting star was born {\it in situ}. An arc-like
structure located  between the star and the denser parts of the cloud to the
north and the north-east at about 18\arcsec~ from the star might represent a
shock due to the  supersonic motion of the star. It is not very clearly visible
on Fig. 1 but is well visible on the red POSS1 image delivered by ALADIN 
(http://aladin.u-strasbg.fr). 

Our CVF observation shows a small source surrounded by a faint extended
emission (Fig. 1). The absolute positioning of the ISOCAM images is uncertain
due to the lens wheel not always returning to its nominal position (the maximum
error is 2 times the Pixel Field of View, viz. 12 arcsec here) and the images
have to be recentered on known objects when possible. Since we see on the
spectra of the pixels corresponding to the strong source a continuum between
5.15 and 5.5\,\mum~ that we attribute to the photospheric emission of the
exciting star (see Fig. 3), we have recentered the CVF image at these
wavelengths on the nominal SIMBAD  position of BD +69\degrees1231
($\alpha$(J2000) = 22h 13m 27s, $\delta$(J2000) =
70\degrees~15\arcmin~18\arcsec). This resulted in Fig. 1. Unfortunately the 
astrometry of BD +69\degrees1231 is itself somewhat problematic  due to the
surrounding bright nebulosity and the location of the CVF image on the optical
image is correspondingly uncertain. Also, the proper motion of the star from
the original BD position is unknown.

Fig. 2 presents a set of CO(2--1) spectra of the molecular cloud obtained by us
at the Caltech Submillimeter 
Observatory with 30\arcsec~ HPBW and 30\arcsec~ sampling. These observations
are better sampled than the CO(2--1) map of Kemper et al. (\cite{Kemper};
21\arcsec~ HPBW, 60\arcsec~ sampling) but are in good agreement. The line
intensity peaks in the direction of the star, probably due to local
excitation of CO. The higher--resolution $^{13}$CO(2--1) map of Kemper et al. 
(\cite{Kemper}; 21\arcsec~ HPBW, 20\arcsec~ sampling) shows that this is not 
the position of maximum column density which peaks well to the NE of the 
star, confirming this conclusion. The $^{12}$CO profiles are double peaked in 
this area,
in particular towards the arc, and this makes difficult the search for line 
wings which would be a signature of a shock. There is however some suggestion
of a negative--velocity wing near the star, in the same sense as the radial 
velocity of this star.

\begin{figure*}[t!]
\vspace{-0.6cm}
{\psfig{file=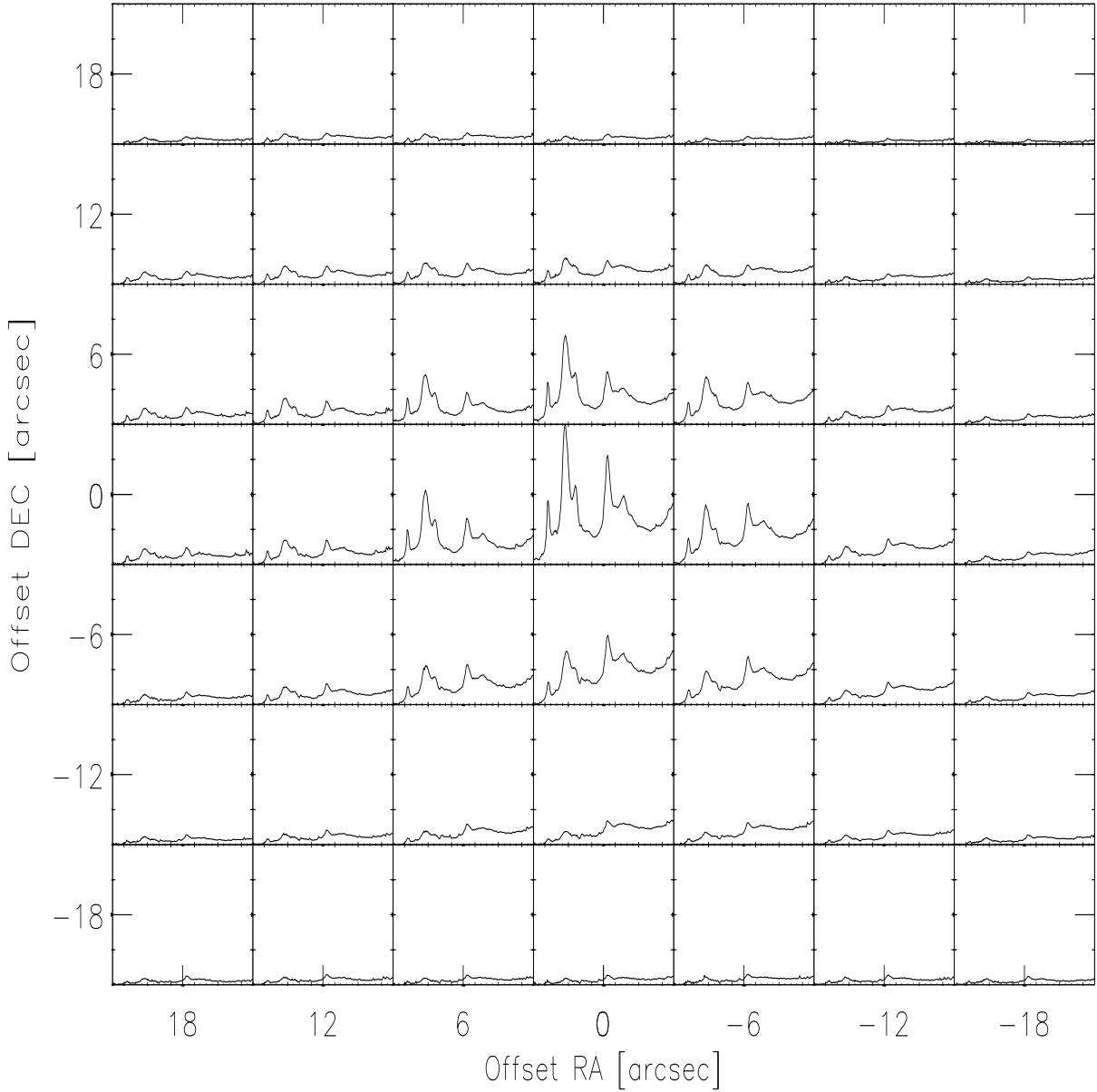,width=18.0cm,height=18.0cm,angle=0.0}}

\vspace{-0.4cm}
\caption{ISOCAM CVF spectra of the central 7$\times$7 CVF spectra of
Ced\,201 on a grid with 6\arcsec$\times$6\arcsec~ spacing. The zodiacal 
light has been subtracted. North is to the top and east is to the left. For
each plot, the spectral range is from 5\,\mum~ to 17\,\mum~ and the flux scale
spans from 0.0 to 11.0 mJy/arcsec$^2$. The offsets are expressed with respect
to the centroid of the total mid-infrared emission, see Fig. 1.}
\end{figure*}

Fig. 3 displays a set of 7$\times$7 CVF spectra on a grid with 
6\arcsec$\times$6\arcsec~ spacing centered on the exciting star. One sees
typical AIB spectra near the emission peak evolving towards fainter, different
spectra 12--18 arcseconds away. The spectra near the peak show not only the
classical AIBs, but also the S(3) rotation line of H$_2$ at 9.6\,\mum~ and the
S(5) line at 6.91\,\mum~ or the line of  [\ArII] at 6.98\,\mum, which cannot be
separated at our resolution. The 12.7\,\mum~ AIB might be contaminated by the
[\NeII] 12.8 \mum~ line. These spectra are typical for a low--excitation
photodissociation region (PDR). The AIBs are superimposed on a continuum rising
towards long  wavelengths. This could be the continuum emission of \mbox{3--D}
very small  grains (VSGs). The AIBs are broader and much fainter relative to
the  continuum. The 11.3\,\mum~ AIB is now the strongest one at this location.
The  7.7 and 8.6\,\mum~ AIBs are merged into a single broad band.  The most
striking feature is the strong emission plateau extending from 11 to 14\,\mum.
These spectra resemble Class B spectra (Tokunaga  \cite{Tokunaga97}) 
(Class A spectra are the usual AIB spectra).

\section{Discussion and conclusions}

Fig. 3 shows a behaviour of the spectrum with radiation field similar
to that observed by Uchida et al. (\cite{Uchida00}) in some other
reflection nebulae.  We find a trend for the 7.7\,\mum~AIBs to become
broader at {\it fainter} radiation
fields away from the exciting star, Fig. 4. The trend is not clear for the 
other bands, but we should not forget that there is contamination by the
AIBs from foreground and background material. We suggests that some
carbonaceous material that emits the continuum and broad bands far from the
star is processed through the effect of the star 
that moves through the molecular cloud, producing AIB carriers: the 
continuum is only 2 times fainter 12\arcsec~ from
the star than close to the star, demonstrating the partial disappearance
of its carriers near the star while the AIBs become very strong. The very
appearance at relatively large distances from the star (at least 18\arcsec) 
of a continuum rising towards long wavelengths is surprising.

This continuum must be emitted by VSGs 
heated by single (visible) photons. These grains must be quite smaller than
the ``classical'' VSGs which require a strong radiation field to emit
in the wavelength range of ISOCAM (Contursi et al. \cite{Contursi}).
The spectra seen far from the star remind strongly of Class B spectra
seen in carbon--rich 
proto--planetary nebulae (Guillois et al. \cite{Guillois96}). In the 
laboratory, such spectra are produced (in absorption) by natural coals
rich in aromatic cycles (Guillois et al. \cite{Guillois96}) or by a-C:H
materials produced by laser pyrolysis of hydrocarbons (Herlin et al. 
\cite{Herlin98}; Schnaiter et al. \cite{Schnaiter99}). In Ced\,201,
these carbonaceous grains must be very small (radius of the order of 1 nm)
since they are heated transiently by visible photons to the temperatures
of $\simeq$ 250 K necessary to emit the observed mid--IR features. 
Since this is also the approximate size of the classical AIB carriers
it would not be appropriate to say that these particles release these carriers.
One should invoke instead chemi--physical transformations like aromatization
(Ryter \cite{Ryter}). In any case, these
small grains were already present before the star penetrated the molecular 
cloud. They are expected to be present elsewhere in the ISM.

\begin{figure}[t!]
\vspace{-0.6cm}
\psfig{file=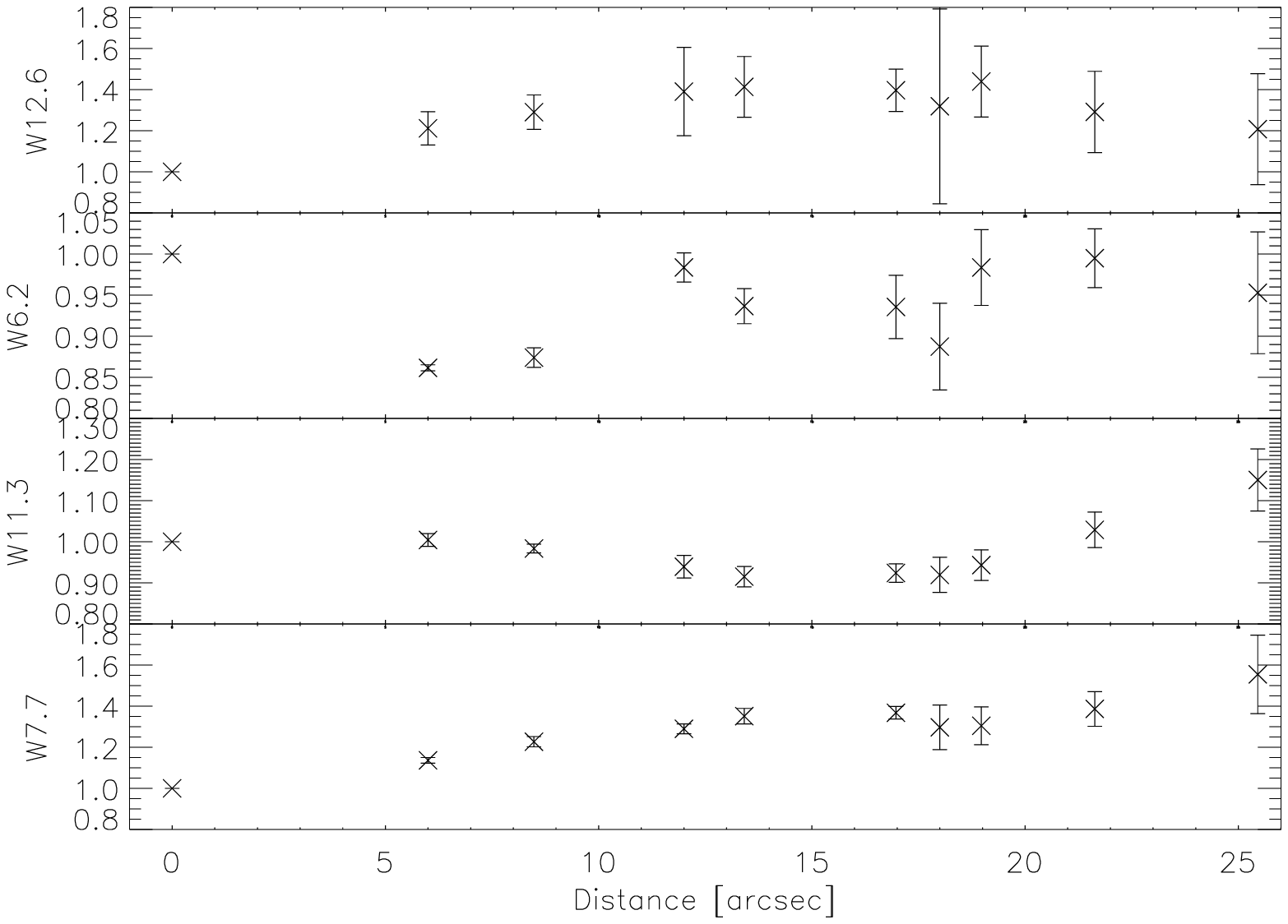,width=8.8cm,height=8.8cm,angle=0.0} 
\caption{Lorentzian widths of four AIB features from Fig. 3 plotted as a
function of the angular distance to the exciting star, normalized to the width
of the Lorentzian at zero distance. The error bars show the internal
dispersion, i.e. the dispersion of values obtained at the same distance from
the central star (i.e. 1 value at distance 0 -- hence null dispersion, 4 values
at distance 6 arcsec, etc.)}
\end{figure}

We find that assuming spherical symmetry the total emissivity per unit  volume
in the CVF spectral range (5.15 to 17\,\mum) decreases approximately  as the
inverse square root of the distance to the star, independently of the shape of
the spectrum. We base the following calculation on the light diffusion model of
Witt et al.  (\cite{Witt}), who postulate a uniform density that we find equal
to n(H) = 2n(H$_2$) = 1800 cm$^{-3}$. This density corresponds to a visual 
extinction of only 0.02 mag. per 6\arcsec~ angular distance. Since the 
integrated mid--IR emission is proportional to the stellar  radiation density,
we can derive an energy conversion efficiency, $\eta$ = 7.5 \%~ (emitted in
4$\pi$ steradians). Knowing
the density, the  radiation field of the B9.5\,V star, and adopting an
absorption cross--section  per carbon atom of $3\times 10^{-18}$ cm$^{-2}$
(Allamandola et al.  \cite{Allamandola89}) and an interstellar carbon abundance
C/H=$2.3\times 10^{-4}$ (Snow \& Witt \cite{Snow95};  see also Andrievsky et al.
\cite{Andrievsky} for carbon abundance in B stars), we find that the fraction
of carbon locked in the very small particles emitting in the mid--IR is 15\% .
 This is in agreement with the independent analysis of Kemper et al.
(\cite{Kemper}) who find that 10\%  of the gas--phase carbon is locked in
particles  smaller than 1.5 nm. The {\it absorption} by these particles
contributes  to 1/4 of the {\it extinction} in the visible, a considerable
fraction. 

We have examined the ratio of the strength of the 6.2\mum~ (a vibration of
the aromatic skeleton) and of the
11.3\mum~AIB  features (a C--H bending mode) as a function of distance to the 
exciting star. No variation has been seen in this ratio within our rather 
large (30 \%) errors.

It is interesting to note that Witt et al. (\cite{Witt}) find from UV--visible
scattering studies of Ced\,201 that the grains responsible for the visible
scattering have a narrow size distribution skewed towards big, 
wavelength--sized grains. This is mostly seen within about 20\arcsec~ from 
the star. On the other hand the very strong 2175 \AA~ extinction band suggests
the presence of many smaller carbon grains. All this is evidence for
grain processing by the radiation of the star, or/and by
shattering of grains by a shock wave associated with its motion. The very
small carbonaceous grains we see in Ced\,201 through their mid--IR emission
might be responsible for the 2175 \AA~ band.
 
In summary, we have shown the existence of transformations near a star of very
small, \mbox{3--D} hydrogenated carbonaceous grains which are probably  present
everywhere in the ISM. This transformation of carbonaceous grains into AIB
carriers might be due either to the radiation of a B9.5V star, or to a shock
induced by the supersonic  motion of the star through the molecular cloud.
Although the  existence of a shock is indicated by an optical arc visible
18\arcsec~ from the star (and also suspected from our CO observations), there
is no  obvious sign of it in the spectra shown in Fig. 3 and it may
not be the cause of the transformation. The very small grains as well as the
AIB carriers formed  from them are excited by the light of the star. The
conversion efficiency between received excitation energy and mid--IR emitted 
energy is
of the order of 7.5\%  for both kinds of particles, and the fraction of carbon
locked in these particles is approximately 15\%.
\begin{acknowledgements} We thank Cecile Reynaud and Olivier Guillois for 
interesting discussions, Kris Sellgren for suggesting Fig. 4  and the referee
Adolf N. Witt for his useful remarks.  The CSO is funded by NSF contract AST
96-15025.
\end{acknowledgements}

\begin{thebibliography}{}




\bibitem[1985]{Allamandola85}
Allamandola L.J., Tielens A.G.G.M., Barker J.R. 1985, ApJ 290, L25

\bibitem[1989]{Allamandola89}
Allamandola L.J., Tielens A.G.G.M., Barker J.R. 1989, ApJS 71, 733

\bibitem[1999]{Andrievsky}
Andrievsky S.M., Korotin S.A., Luck R.E., Kostynchuk L. Yu. 1999, A\&A 350, 598



\bibitem[1990]{Boulanger}
Boulanger F., Falgarone, E., Puget J.L., Helou G. 1990, ApJ 364, 136

\bibitem[1991]{Casey}
Casey S.C. 1991, ApJ 371, 183

\bibitem[1996a]{CCesarsky}
Cesarsky C.J., Abergel A, Agn\`ese P. \etal ~1996a, A\&A 315, L32


\bibitem[1996b]{M17}
Cesarsky D., Lequeux J., Abergel A. et al., 1996b, A\&A 315, L309


\bibitem[1992]{Cherchneff}
Cherchneff I., Barker J.R., Tielens A.G.G.M. 1992, ApJ 401, 269


\bibitem[1998]{Contursi}
Contursi A., Lequeux J., Hanus M., et al. 1998, A\&A 336, 662

\bibitem[1999]{Coulais}
Coulais A., Abergel A. 1999, in {\it The Universe as seen
by ISO}, ed. P. Cox \& M.F. Kessler, ESA SP-427, 61

\bibitem[1998]{Cox}
Cox P., Boulanger F., Huggins P.J., et al. 1998, ApJ 495, L23

\bibitem[1993]{Duley93}
Duley W.W., Jones A.P., Taylor S.D., Williams D.A. 1993, MNRAS 260, 415

\bibitem[1988]{Duley88}
Duley W.W., Williams D.A. 1988, MNRAS 231, 969



\bibitem[1998]{Gry}
Gry C., Boulanger F., Falgarone E. et al., 1998, A\&A 331, 1070

\bibitem[1996]{Guillois96}
Guillois O., Nenner I., Papoulard C., Reynaud C. 1996, ApJ 464, 810

\bibitem[1998]{Herlin98}
Herlin N., Bohn I., Reynaud C. et al., 1998, A\&A 30, 1127



\bibitem[1999]{Kemper}
Kemper C., Spaans M., Jansen D.J. et al., 1999, ApJ 515, 649


\bibitem[1984]{Leger}
L\'eger A., Puget J.-L. 1984, A\&A 137, L5


\bibitem[1999]{Pagani99}
Pagani L., Lequeux J., Cesarsky D. et al., 1999, A\&A 351, 447




\bibitem[1991]{Ryter}
Ryter C. 1991, Annales de Physique 16, 507


\bibitem[1999]{Schnaiter99}
Schnaiter M., Henning, Th., Mutschke H. et al., 1999, ApJ 519, 687 


\bibitem[1997]{Scott97}
Scott A.D., Duley W.W., Pinho G.P. 1997, ApJ 489, L193

\bibitem[1983]{Sellgren83}
Sellgren K, Werner M.W., Dinerstein H.L. 1983, ApJ 271, L13

\bibitem[1989]{Snow95}
Snow T.P., Witt A.N. 1995, Science 270, 1455

\bibitem[1999]{Starck}
Starck J.L., Abergel A., Aussel H. et al., 1999, A\&AS 134, 135

\bibitem[1997]{Tokunaga97}
Tokunaga A. 1997, in ASP Conf. Ser. 124, Diffuse Infrared Radiation and the
IRTS, ed. H. Okuda, T. Matsumoto, T.L. Roellig, San Francisco (ASP), 149 

\bibitem[1998]{Uchida98}
Uchida K.I., Sellgren K., Werner M. 1998, ApJ 493, L109

\bibitem[2000]{Uchida00}
Uchida K.I., Sellgren K., Werner M.W., Houdashelt M.L. 2000, ApJ in press
(astro-ph/9910123)



\bibitem[1987]{Witt}
Witt, A.N., Bohlin, R.C., Stecher, T.P., Graff, S.M. 1987, ApJ 321, 912

\end{thebibliography}
\end{document}